\documentclass{statsoc}

\usepackage[a4paper,margin=1in]{geometry}
\usepackage{etoolbox}

\makeatletter
\patchcmd{\@makecaption}
  {\parbox}
  {\advance\@tempdima-\fontdimen2} 
  {}{}
\makeatother  
\usepackage[dvipdfmx]{graphicx} 
\usepackage{bmpsize}
\usepackage{graphicx}
\usepackage[textwidth=8em,textsize=small]{todonotes}
\usepackage{amsmath}
\usepackage{natbib}
\usepackage{hyperref}
\usepackage{url}
\usepackage{graphicx}
\usepackage{amsmath, amssymb}
\usepackage{bbold}
\usepackage{mathbbol}
\usepackage{float}
\usepackage{xcolor}
\usepackage{bm} 
\usepackage{setspace}
\usepackage{multirow}
\usepackage{lscape}
\usepackage{rotating}
\usepackage{adjustbox}
\usepackage{booktabs}
\usepackage{ragged2e}
\usepackage[caption = false]{subfig} 
\usepackage{float}
\def\*#1{\mathbf{#1}}
\def\^#1{\amsmathbb{#1}}
\def\##1{\mathbb{#1}}
\DeclareSymbolFontAlphabet{\amsmathbb}{AMSb}%

\title[Variable Selection in Functional Linear Concurrent Regression]{Variable Selection in Functional Linear Concurrent \\Regression}
\author[R. Ghosal, A. Maity, T. Clark and S.B. Longo]{Rahul Ghosal$^1$, Arnab Maity$^1$, Timothy Clark$^2$ and Stefano B Longo$^2$}
\address{$^{1}$ North Carolina State University (Department of Staistics)}
\address{$^{2}$ North Carolina State University (Department of Sociology and Anthropology)}

\begin{document}
\begin{singlespace}
\begin{abstract}
We propose a novel method for variable selection in functional linear concurrent regression. Our research is motivated by a fisheries footprint study where the goal is to identify important time-varying socio-structural drivers influencing patterns of seafood consumption, and hence fisheries footprint, over time, as well as estimating their dynamic effects. We develop a variable selection method in functional linear concurrent regression extending the classically used scalar on scalar variable selection methods like LASSO, SCAD, and MCP. We show in functional linear concurrent regression the variable selection problem can be addressed as a group LASSO, and their natural extension; group SCAD or a group MCP problem. \hspace{- 1.3 mm}Through simulations, we illustrate our method, particularly with group SCAD or group MCP penalty, can pick out the relevant variables with high accuracy and has minuscule false positive and false negative rate even when data is observed sparsely, is contaminated with noise and the error process is highly non-stationary. We also demonstrate two real data applications of our method in studies of dietary calcium absorption and fisheries footprint in the selection of influential time-varying covariates.
\end{abstract}
\end{singlespace}
\vspace*{- 3 mm}

\section{Introduction}
\label{sec:intro}
Function on function regression is an active area of research in functional data with new statistical methods frequently emerging to address data where both the response variable and the covariates are functions over some continuous index such as time. The functional concurrent regression model is a special case of function on function regression where the predictor variables influence the response variable only through their value at the current time point \citep{kim2016general}. The commonly used functional linear concurrent regression model assumes a linear relationship between the response and the predictors, where the value of the response at a particular time point is modelled as a linear combination of the covariates at that specific time point, and the coefficients of the functional covariates are univariate smooth functions over time \citep{Ramsay05functionaldata}. Multiple methods exist in literature for estimation of these regression functions in functional linear concurrent regression and the closely related varying coefficient model \citep{hastie1993varying}, using kernel-local polynomial smoothing \citep{wu1998asymptotic, hoover1998nonparametric,fan1999statistical, kauermann1999model}, polynomial spline \citep{huang2002varying,huang2004polynomial}, smoothing spline \citep{hastie1993varying,hoover1998nonparametric,chiang2001smoothing,eubank2004smoothing} among many others. Similar to classical scalar regression, when there are a large number of covariates present, the primary interest might be to select only the set influential variables and estimate their effects. While doing significance testing and building confidence bands can help for assessing the individual effect of a predictor, they are computationally infeasible when the number of covariates is large. Thus arises the need to perform variable selection in functional linear concurrent regression.

Our research in this article is motivated by a fisheries footprint study where the goal is to identify important time-varying socio-structural and economic drivers influencing fisheries footprint (Global Footprint Network's measure of the total marine area required to produce the amount of seafood products a nation consumes) and to estimate their time-varying effects. Although, a number of variable selection methods have been developed for scalar on function regression \citep{gertheiss2013variable,fan2015functional} and function on scalar regression \citep{chen2016variable}, literature for variable selection in functional linear concurrent regression is relatively sparse. Recently \cite{goldsmith2017variable} developed a variable selection method for functional linear concurrent model using a variational Bayes approach with sparsity being introduced through a spike and slab prior on the coefficients of the basis expansion of the regression functions. In this article, we propose a variable selection method in functional linear concurrent regression extending the classically used variable selection methods like LASSO \citep{tibshirani1996regression}, SCAD \citep{fan2001variable}, and MCP \citep{zhang2010nearly}.

Our work is inspired by \cite{gertheiss2013variable}, where they show the variable selection problem in scalar on function regression scenario can be reduced to a group LASSO \citep{yuan2006model} problem. We have shown in functional linear concurrent regression the variable selection problem can be addressed as a group LASSO, and their natural extension group SCAD or group MCP problem. \cite{chen2016variable} also used group MCP for their variable selection in function on scalar regression. Our model is fundamentally different from them in the sense the covariates we consider are time-varying functions and possibly observed with measurement error. Our method is similar to \cite{wang2008variable} in which they use a group SCAD penalty for variable selection in varying coefficient models, but we propose a different penalty on the coefficient functions which simultaneously penalizes departure from sparsity as well as roughness of the coefficient functions, and our research shows there is much to be gained by using the group MCP penalty. We employ a pre-whitening procedure similar to \cite{chen2016variable} to take into account the possible temporal dependence present within functions. We also consider that the covariates might be contaminated with measurement error and therefore use functional principal component analysis (FPCA) to get denoised trajectories of the covariates, which improves the estimation accuracy of our approach. Through simulations, we illustrate the proposed method, particularly with group SCAD or group MCP penalty, can pick out the relevant variables with very high accuracy and has minuscule false positive and false negative rates even when data is observed sparsely and is contaminated with measurement error. We demonstrate two real data applications of our method in the study of dietary calcium absorption \citep{davis2002statistical} and the fisheries footprint study in selection of the influential time-varying covariates.

The rest of the article is organized as follows. In Section \ref{sec:method}, we present our modelling framework and illustrate our variable selection method. In Section \ref{sec:sim_stud}, we conduct a simulation study to evaluate the performance of our method and summarize the simulation results. In Section \ref{sec:real_dat}, we go back to the two real data examples; calcium absorption study and fisheries footprint study, apply our variable selection method to find out the influential covariates and present our findings. We conclude in Section \ref{con} with a discussion about some limitations and possible extensions of our work.
\vspace*{- 6 mm}
\section{Methodology}
\label{sec:method}
\vspace*{- 3 mm}
\subsection{Modelling Framework and Variable Selection Method}
\label{sec:mf}
Suppose that the observed data for the $i$-th subject 
is $\{Y_i(t), Z_{i1}(t),Z_{i2}(t),\ldots,Z_{ip}(t)\}$  ($i=1,2,\ldots,n$), where $Y_i(\cdot)$ is a functional response and $Z_{i1}(\cdot)$,$Z_{i2}(\cdot)$,$\ldots,Z_{ip}(\cdot)$ are the
corresponding functional covariates. We assume the covariates and the response are observed on a fine and regular grid of points $S= \{t_{1},t_{2},\ldots,t_{m} \} \subset S=[0,T] $ for some $T>0$, and the covariates are measured without any error. We discuss later in this section how our model and the proposed method can be easily extended to accommodate more general scenarios where the covariates are contaminated with measurement error and observed sparsely. We consider a functional linear concurrent regression model of the form,
\begin{equation}
Y_i(t)=\sum_{j=1}^{p}Z_{ij}(t)\beta_{j}(t)+\epsilon_{i}(t),
\end{equation}
where $\beta_{j}(t)$ ($j=1,2,\ldots,p$) are smooth functions (with finite second derivative) representing the functional regression parameters. We assume $Z_{ij}(\cdot)$ are independent and identically distributed (i.i.d.) \hspace{- 0.8mm}copies of ${Z}_j(\cdot)$ ($j=1,2,\ldots,p$), where ${Z}_{j}(\cdot)$ s are underlying smooth stochastic processes. We further assume $\epsilon_{i}(\cdot)$ are i.i.d copies of $\epsilon(\cdot)$, which is a mean zero stochastic process. The model (1) in stacked form can be rewritten as $Y(t)=Z(t)\beta(t)+\epsilon(t)$. 
 Generally in functional linear concurrent regression, estimation is done \citep{Ramsay05functionaldata} by minimizing the penalized residual sum of square, $ SSE(\beta)=\int r(t)^{T}r(t)dt + \sum_{j=1}^{p}\lambda_j\int (L_j\beta_j(t))^2dt$, where $r(t)=Y(t)-Z(t)\beta(t)$. For example when $L_j=I$, we minimize $\int r(t)^{T}r(t)dt + \sum_{j=1}^{p}\lambda_j\int (\beta_j(t))^2dt$.
 Now suppose $\{\theta_{kj}(t),  k=1,2,\ldots,k_j\}$ is a set of known basis functions for $j=1,2,\ldots,p$ . We model the unknown coefficient functions using basis function expansion as $\beta_{j}(t)=\sum_{k=1}^{k_{j}} b_{kj}\theta_{kj}(t)=\bm\theta_{j}(t)^{T}\*b_{j}$, where $\bm\theta_{j}(t)=[\theta_{1j}(t),\theta_{2j}(t),\ldots ,\theta_{K_jj }(t)]^T$ and $\*b_{j}=(b_{1j},b_{2j},\ldots ,b_{k_j j})^T$ is a vector of unknown coefficients. In this article, we use B-spline basis functions, however, other basis functions can be used as well. Then the minimization in the example mentioned above can be carried out by minimizing
 $\int \{Y(t)-Z(t)\bm\Theta(t)\*b\}^{T}\{Y(t)-Z(t)\bm\Theta(t)\*b\}dt +\*b^{T}\^R\*b$. Here $\*b$, $\bm\Theta(t)$ and penalty matrix $\^R$ are defined in stacked form as  
$\*b= (\*b_1^T,\*b_2^T,\ldots,\*b_p^T)^T$, $\bm\Theta(t)=\{\bm\theta_{1}(t)^{T},\bm\theta_{2}(t)^{T},\ldots,\bm\theta_{p}(t)^{T}\}$ and $\^R=diag(\^R_1,\^R_2,\ldots,\^R_p)$, where $\^R_j= \lambda_j\*b_{j}^{T}\{\int\bm \theta_{j}(t)\bm\theta_{j}(t)^{T}dt\}\*b_{j}$. For our variable selection method, we define penalty on the regression functions $\beta_j(\cdot)$ as,
$
P_{\lambda,\psi}\{\beta_j(\cdot)\}=\lambda[\int\beta_j(t)^2dt+\psi\int\{\beta_j^{''}(t)\}^2dt]^{1/2} =\lambda\left(\*b_j^{T}\^R_j\*b_j+\psi \*b_j^{T}\^Q_j\*b_j\right)^{1/2} 
\\=\lambda\left(\*b_j^{T}\^K_{\psi,j}\*b_j\right)^{1/2},$ where $\^K_{\psi,j}=\^R_j+\psi\^ Q_j$, $\^R_j=\{\int\bm\theta_{j}(t)\bm\theta_{j}(t)^{T}dt\}$,  $\^Q_j=\{\int\bm\theta_{j}^{''}(t)\bm\theta_{j}^{''}(t)^{T}dt\}$. This penalty was originally proposed by \cite{meier2009high} and later used by \cite{gertheiss2013variable} for their variable selection method in scalar on function regression. The parameter $\psi\geq0$ controls the amount of penalization on the roughness penalty. The proposed penalty simultaneously penalizes departure from sparsity and roughness of the coefficient functions ensuring the resulting coefficient functions are smooth and small coefficient functions are shrunk to zero introducing sparsity. Subsequently, we propose to minimize the following penalized mean sum of squares of the residuals for performing variable selection,
\begin{equation}
L(\*b)=1/n\int \{Y(t)-Z(t)\bm\Theta(t)\*b\}^{T}\{Y(t)-Z(t)\bm\Theta(t)\*b\}dt +\lambda\sum_{j=1}^{p}(\*b_j^{T}\^K_{\phi,j}\*b_j)^{1/2}.\end{equation}
Since we assume data is observed on a dense equispaced grid, the variable selection in practice is carried out by minimizing the following equivalent criterion,
\begin{equation}
\sum_{i=1}^{n}\sum_{l=1}^{m}[Y_i(t_{l})-\sum_{j=1}^{p}Z_{ij}(t_l)\{\sum_{k=1}^{k_{j}} b_{kj}\theta_{kj}(t_l)\}]^{2}+\lambda mn\sum_{j=1}^{p}(\*b_j^{T}\^K_{\psi,j}\*b_j)^{1/2}.
\end{equation}
Now using Cholesky decomposition of $\^K_{\psi,j}=\^L_{\psi,j}\^L_{\psi,j}^T$ and denoting $\bm\gamma_j=\^L_{\psi,j}^T \*b_j$, the penalized sum of square of residuals can be reformulated as,
\vspace*{- 3 mm}
\begin{align*}
R(\bm{\gamma})&=\sum_{i=1}^{n}\sum_{l=1}^{m}[Y_i(t_{l})-\sum_{j=1}^{p}Z_{ij}(t_l)\{\sum_{k=1}^{k_{j}} b_{kj}\theta_{kj}(t_l)\}]^{2}+\lambda mn\sum_{j=1}^{p}(\*b_j^{T}\^K_{\psi,j}\*b_j)^{1/2}\nonumber\\
&=\sum_{i=1}^{n}\sum_{l=1}^{m}[Y_i(t_{l})-\sum_{j=1}^{p}\*Z_{ij}^*(t_l)^T\*b_{j}]^{2}+\lambda mn\sum_{j=1}^{p}(\*b_j^{T}\^K_{\psi,j}\*b_j)^{1/2}\hspace{2mm} \textit{where $\*Z_{ij}^*(t_l)^T=Z_{ij}(t_l)\times\bm\theta_j(t_l)^T$}\\
&=\sum_{i=1}^{n}\sum_{l=1}^{m}[Y_i(t_{l})-\sum_{j=1}^{p}\tilde{\*Z_{ij}}^*(t_l)^T \bm{\gamma}_{j}]^{2}+\lambda mn\sum_{j=1}^{p}(\bm\gamma_j^{T} \bm\gamma_j)^{1/2} \hspace{2mm} \textit{where $\tilde{\*Z_{ij}}^*(t_l)=\^L_{\psi,j}^{-1}\*Z_{ij}^*(t_l)$}\\
&=\sum_{i=1}^{n}||{\*Y_i}-\^Z_i^*\bm\gamma||_{2}^{2}+\lambda mn\sum_{j=1}^{p}(\bm\gamma_j^{T}\bm\gamma_j)^{1/2}
=\sum_{i=1}^{n}||{\*Y_i}-\sum_{j=1}^{p}\^Z_i^{*j}\bm\gamma_j||_{2}^{2}+\lambda mn\sum_{j=1}^{p}(\bm\gamma_j^{T}\bm\gamma_j)^{1/2},
\end{align*}
where ${\*Y_i}=(Y_i(t_{1}),Y_i(t_{2}),\ldots,Y_i(t_{m}))^{T}$, and $\bm\gamma= (\bm\gamma_1^T,\bm\gamma_2^T,\ldots,\bm\gamma_p^T)^T$ and $\^Z_i^*$  is defined as follows,
\newpage
\begin{align*}
\^Z_i^*&= 
\begin{bmatrix}
    \tilde{\*Z_{i1}}^*(t_1)^T       & \tilde{\*Z_{i2}}^*(t_1)^T & \tilde{\*Z_{i3}}^*(t_1)^T & \dots & \tilde{\*Z_{ip}}^*(t_1)^T \\
    \tilde{\*Z_{i1}}^*(t_2)^T         & \tilde{\*Z_{i2}}^*(t_2)^T    & \tilde{\*Z_{i3}}^*(t_2)^T    & \dots &\tilde{\*Z_{ip}}^*(t_2)^T    \\
    \hdotsfor{5} \\
    \tilde{\*Z_{i1}}^*(t_m)^T         & \tilde{\*Z_{i2}}^*(t_m)^T    & \tilde{\*Z_{i3}}^*(t_m)^T    & \dots & \tilde{\*Z_{ip}}^*(t_m)^T   
\end{bmatrix},
\end{align*} where $\^Z_i^{*j}$ refers to the  $j$th block column in
this matrix. We recognize this minimization problem as performing a
group LASSO \citep{yuan2006model}, where the grouping is introduced
by covariates. In particular, we obtain estimates of $\bm\gamma_j$ by minimizing similar penalized least square as in group LASSO namely;
\begin{align}
\bm\hat{\bm\gamma}&=\underset{\bm\gamma_j,j=1,2,\ldots,p}{\text{argmin}} \hspace{2 mm} \sum_{i=1}^{n}||{\*Y_i}-\sum_{j=1}^{p}\^Z_i^{*j}\bm\gamma_j||_{2}^{2}+\lambda mn\sum_{j=1}^{p}(\bm\gamma_j^{T}\bm\gamma_j)^{1/2}\nonumber\\
&=\underset{\bm\gamma_j,j=1,2,\ldots,p}{\text{argmin}}\hspace{2 mm} \sum_{i=1}^{n}||{\*Y_i}-\sum_{j=1}^{p}\^Z_i^{*j}\bm\gamma_j||_{2}^{2}+\lambda mn\sum_{j=1}^{p}||\bm\gamma_j||_2\nonumber\\
&=\underset{\bm\gamma_j,,j=1,2,\ldots,p}{\text{argmin}}\hspace{2 mm}  \sum_{i=1}^{n}||{\*Y_i}-\sum_{j=1}^{p}\^Z_i^{*j}\bm\gamma_j||_{2}^{2}+mn\sum_{j=1}^{p}P_{LASSO,\lambda}(||\bm\gamma_j||_2) .\end{align}
We extend this group LASSO formulation to non-convex penalties, which are known \citep{breheny2015group,mazumder2011sparsenet} to produce sparser solutions especially when there are large number of variables. In particular, we propose to use two non convex penalties; SCAD \citep{fan2001variable} and MCP \citep{zhang2010nearly}. These two penalties overcome the high bias problem of LASSO as they relax the rate of penalization as the magnitude of the coefficient gets large. SCAD and MCP have been shown to ensure selection consistency and estimation consistency under standard assumptions in the scalar regression case. They also enjoy the so-called oracle property in which they behave like oracle MLE asymptotically. Unlike adaptive LASSO, these methods do not require initial estimates of weights. These facts motivate us to use them in our functional variable selection context. Then the problem of variable selection reduces to a group SCAD or group MCP problem in our modeling setup as follows.

\hspace*{-.7 cm}
\textbf{Group SCAD Method}\par
In this method, we perform variable selection and obtain estimates of $\bm\gamma$ using a penalized least square criterion as in (4), where we now use
a group SCAD penalty on the coefficients instead of group LASSO. In particular we estimate,
\begin{equation}
    \hat{\bm\gamma}=\underset{\bm\gamma_j,j=1,2,\ldots,p}{\text{argmin}}\hspace{2 mm}  \sum_{i=1}^{n}||{\*Y_i}-\sum_{j=1}^{p}\^Z_i^{*j}\bm\gamma_j||_{2}^{2}+mn\sum_{j=1}^{p}P_{SCAD,\lambda,\phi}(||\bm\gamma_j||_2),
\end{equation}
where $P_{SCAD,\lambda,\phi}(||\bm\gamma_j||_2)$ is defined in the following way:
\newpage
\begin{equation*}
P_{SCAD,\lambda,\phi}(||\bm\gamma_j||_2)=
\begin{cases}
\lambda||\bm\gamma_j||_2\hspace{4.15 cm} \text{if $||\bm\gamma_j||_2\leq \lambda$}.\\
\frac{\lambda\phi||\bm\gamma_j||_2-.5(||\bm\gamma_j||_2^{2}+\lambda^2)}{
\phi-1} \hspace{2.1 cm} \text{if $\lambda< ||\bm\gamma_j||_2\leq \lambda\phi$}.\\
.5\lambda^2(\phi+1) \hspace{3.45 cm} \text{if $||\bm\gamma_j||_2>\lambda\phi$}.\\
\end{cases}
\end{equation*}
\textbf{Group MCP Method}\par
For Group MCP method we estimate $\bm\gamma$ as \begin{equation}\hat{\bm\gamma}=\underset{\bm\gamma_j,j=1,2,\ldots,p}{\text{argmin}}\hspace{1 mm}\sum_{i=1}^{n}||{\*Y_i}-\sum_{j=1}^{p}\^Z_i^{*j}\bm\gamma_j||_{2}^{2}+mn\sum_{j=1}^{p}P_{MCP,\lambda,\phi}(||\bm\gamma_j||_2),\end{equation}\\
where $P_{MCP,\lambda,\phi}(||\bm\gamma_j||_2)$ is defined as :
\begin{equation*}
P_{MCP,\lambda,\phi}(||\bm\gamma_j||_2)=
\begin{cases}
\lambda||\bm\gamma_j||_2-\frac{||\bm\gamma_j||_2^2}{2\phi}\hspace{1.4 cm} \text{if $||\bm\gamma_j||_2\leq \lambda\phi$}.\\
.5\lambda^2\phi \hspace{3.05 cm} \text{if $||\bm\gamma_j||_2>\lambda\phi$}.\\
\end{cases}
\end{equation*}
\subsection{Incorporating Covariance Structure into Variable Selection}
\label{sec:prewhite}
The variable selection method proposed in Section \ref{sec:mf} does not account for possible correlation in the error process. In reality, however, temporal correlation is more likely to be present within functions. While using an independent working correlation structure can yield consistent and unbiased estimates, incorporating the true covariance structure
in the variable selection criterion (4), (5), or 
(6) may give definite gains in terms of performance, as illustrated by \cite{chen2016variable}. We follow a similar pre-whitening procedure employed by \cite{chen2016variable,kim2016general} to take into account the correct covariance structure. We assume the error process $\epsilon(t)$ has the form $\epsilon(t)=V(t) + w_t$, where $V(t)$ is a smooth mean zero stochastic process with covariance kernel $G(s,t)$ and $w_t$ is a white noise with variance $\sigma^2$. The covariance function of the error process is then given by $\Sigma(s,t)=cov\{\epsilon(s),\epsilon(t)\}=G(s,t) + \sigma^2 I(s=t)$. For data observed on dense and regular grid, the covariance matrix of the residual vector is the given by, $\#\Sigma$=diag \{${\#\Sigma_{m\times m},\#\Sigma_{m\times m},\ldots, \#\Sigma_{m\times m}}$\}, where $\#\Sigma_{m\times m}$ denotes the covariance kernel $\Sigma(s,t)$ evaluated at $S= \{t_{1}, t_{2},\ldots, t_{m} \} $. Now if $\#\Sigma_{m\times m}$ is known, redefining $\*Y_i$ and $\^Z_i^{*j}$ as $\*Y_i=\{\#\Sigma_{m\times m}^{-1/2}\}\*Y_i$, $\^Z_i^{*j}=\{\#\Sigma_{m\times m}^{-1/2}\}\^Z_i^{*j}$, the same penalized criterion (4), (5) or (6) can be used to perform variable selection. 

In reality $\#\Sigma$ is unknown, and we need an estimator $\hat{\#\Sigma}$. In the context of functional data, we want to estimate $\Sigma(\cdot,\cdot)$ nonparametrically. If we had the original residuals $\epsilon_{ij}$ available, we could use functional principal component analysis (FPCA), e.g., \cite{yao2005functional2} or \cite{zhang2007statistical} to estimate $\Sigma(s,t)$. If the covariance kernel $G(s,t)$ of the smooth part $V(t)$ is a Mercer kernel \citep{j1909xvi}, by Mercer's theorem $G(s,t)$ must have a spectral decomposition
$$G(s,t)=\sum_{k=1}^{\infty}\lambda_k\phi_k(s)\phi_k(t),$$
where $\lambda_1\geq\lambda_2\geq \ldots0$ are the ordered eigenvalues and $\phi_k(\cdot)$s are the corresponding eigenfunctions. Thus we have the decomposition $\Sigma(s,t)=\sum_{k=1}^{\infty}\lambda_k\phi_k(s)\phi_k(t) + \sigma^2 I(s=t)$. Given $\epsilon_{t_{ij}}=V(t_{ij}) + w_{ij}$, one could employ FPCA based methods to get $\hat{\phi}_k(\cdot)$, $\hat{\lambda}_k$s and $\hat{\sigma^2}$. So an estimator of $\Sigma(s,t)$ can be formed as
$$\hat{\Sigma}(s,t)=\sum_{k=1}^{K}\hat{\lambda_k}\hat{\phi_k}(s)\hat{\phi_k}(t) + \hat{\sigma^2} I(s=t),$$ 
where $K$ is large enough for the convergence to hold and is typically chosen such that percent of variance explained (PVE) by the selected eigencomponents exceeds some pre-specified value such as $99\%$ or $95\%$. In reality, we don't have  the original residuals $\epsilon_{ij}$ and use the full model (1) to obtain residuals $e_{ij}=Y_{i}(t_j)-\hat{Y_{i}}(t_j)$. Then treating $e_{ij}$ as our original residuals, we obtain $\hat{\Sigma}(s,t)$ using FPCA. 

\hspace*{- 7 mm}
\textit{Remark 1:}
We use cubic B-spline basis with the same number of basis functions to model the regression functions $\beta_j(t)$s, where the number of basis is large so the basis is rich enough. For selection of the tuning parameter $\psi$ (for smoothness) and the penalty parameter $\lambda$, we use the Extended Bayesian information criteria (EBIC) \citep{chen2008extended} corresponding to the equivalent linear model of criterion (4), (5) or (6) and this has shown good performance in our simulation study. \cite{chen2008extended} established consistency of EBIC under standard assumptions and illustrated its superiority over other methods like cross-validation, AIC, and BIC, which tend to over select the variables. For tuning parameter $\phi$ we use the values $4$ for SCAD and $3$ for MCP, as proposed by the original authors. For model fitting we use `grpreg' package \citep{breheny2019package} in R.

\hspace*{- 7 mm}
\textit{Remark 2:}
In practice, we recommend standardizing the variables either using Euclidean norm (automatically performed in `grpreg') or using FPCA based methods ($X^{*}_j(t)=\frac{X_j(t)-\mu_j(t)}{\sigma_j(t)}$), which is especially useful for highly sparse data where some B-splines might not have observed data on its support. This can help in faster convergence of the proposed method. We performed both the standardization methods in our simulation studies and obtained very similar results.

\subsection{Extension to Sparse data and Noisy Covariates}
\label{sec:extension}
More generally we can consider the case where data is observed sparsely and covariates are observed with
measurement error. This is most often the case for longitudinal data. Here the observed data is the response \{($Y_i(t_{ij}),t_{ij}),
j=1,2,\ldots,m_i$\} and the observed covariates \{($U_1(t_{1ij}),t_{1ij}),
j=1,2,\ldots,m_{1i}$\},
\{($U_2(t_{2ij}),t_{2ij}),
j=1,2,\ldots,m_{2i}$\},\ldots,\{($U_p(t_{pij}),t_{pij}), \hspace{1mm}j=1,2,\ldots,m_{pi}$\}. Let us denote $U_k(t_{kij})$s, ($k=1,2,3,\ldots,
p$) by $U_{ijk}$. Here $U_{ijk}$ s represent the observed
covariates with measurement error, i.e., we have 
$U_{ijk}=Z_k(t_{kij})+e_{ijk}$ for $i=1,2,\ldots,n$, $j=1,2,\ldots,m_{ki}$
and $k=1,2,\ldots,p$. The measurement error $e_{ijk}$ are assumed to be white noises with zero mean and variance
$\sigma_k^2$. In sparse data set up it is generally assumed \citep{kim2016general} although
individual number of observations $m_{i}$ is small,
$\bigcup_{i=1}^{n}\bigcup_{j=1}^{m_i}{t_{ij}}$ is dense in $[0,T]$.
Then we reconstruct the original curves from the observed sparse
and noisy curves using FPCA methods \citep{yao2005functional2} by estimating the eigenvalues and
eigenfunctions corresponding to the original curves. \cite{li2010uniform} proved uniform convergence of
the mean, eigenvalues and eigenfunctions associated with the curves
for both dense and in particular sparse design under suitable
regularity conditions. For prediction of the scores, we use PACE
method as in \cite{yao2005functional2}. Then these estimates are put together using  Karhunen-Lo$\grave{e}$ve expansion
(Karhunen, Loeve 1946) to get estimates $\hat{Z}_{ik}(\cdot)$  of the true curves $Z_{ik}(\cdot)$ as
$\hat{Z}_{ik}(t)=\hat{\mu}_k(t)+\sum_{s=1}^{S}\hat{\zeta}_{isk}\hat{
\psi}_{sk}(t)$, where the number of eigenfunctions $S$ to use is chosen using the percent of variance explained (PVE) criterion, which is the percentage of variance explained by the first few eigencomponents. Alternatively one can also use multivariate FPCA \citep{doi:10.1080/01621459.2016.1273115} instead of running FPCA on each predictor variable separately. Then for sparse data observed on irregular grid and
observed with measurement error, we use
\{$Y_i(t_{ij}),\hat{Z}_{i1}(t_{ij}),\hat{Z}_{i2}(t_{ij}),\ldots,\hat
{Z}_{ip}(t_{ij})  j=1,2,\ldots,m_i\}_{i=1}^{n}$ as our original data for performing variable selection.
\section{Simulation Study}
\label{sec:sim_stud}

\subsection{Simulation Setup}
In this section, we evaluate the performance of our variable selection method using a simulation study. To this end we generate data from the model,
$$Y_i(t)=\beta_0(t)+\sum_{j=1}^{20}Z_{ij}(t)\beta_{j}(t)+\epsilon_{i}(t),\hspace{2mm} i=1,2,\ldots,n, \hspace{2mm} t\in [0,100].$$
The regression functions are given by $\beta_{0}(t)= 8sin(\pi t/50)$, $\beta_{1}(t)= 5sin(\pi t/100)$, $\beta_{2}(t)= 4sin(\pi t/50)+4cos(\pi t/50)$, $\beta_{3}(t)=25e^{-t/20}$ and rest of the 
$\beta_j(t)=0$ for $j=4,5,6,\ldots,20$, i.e., the last 17 covariates are not relevant. The original covariates
$Z_{ij}(\cdot)\stackrel{iid}{\sim} Z_j(\cdot)$, where $Z_j(t)$ ($j=1,2,\ldots,20$) are given by $Z_j(t)=a_j\hspace{1mm}\sqrt[]{2}sin(\pi j t/400)+ b_j\hspace{1mm}\sqrt[]{2}cos(\pi j t/400)$, where $a_j\sim \mathcal{N}(50,2^2)$, $b_j\sim \mathcal{N}(50, 2^2)$. We  moreover assume that $Z_{ij}(t)$ are observed with measurement error i.e., we observe $U_{ij}(t)=Z_{ij}(t)+\delta_j$, where $\delta_j\sim \mathcal{N}(0, 0.6^2)$. The error process $\epsilon_{i}(t)$ is generated as follows;
$$\epsilon_i(t)=\xi_{i1} cos(t) +\xi_{i2}sin(t) + N(0,1),$$
where $\xi_{i1}\stackrel{iid}{\sim}\mathcal{N}(0, 0.5^2)$  and $\xi_{i2}\stackrel{iid}{\sim}\mathcal{N}(0,0.75^2)$. The response $Y_i(t)$ and noisy covariate $U_{ij}(t)$'s are observed sparsely for randomly chosen $m_{i}$ points in $S$, the set of $m = 81$ equidistant time points in $[0,100]$ and $m_{i}\stackrel{iid}{\sim} Unif\{30,31,\ldots,41\}$.
Three sample sizes $n\in\{100,200,400\}$ are considered. For each sample size, we use 500 generated datasets for evaluation of our method.

\subsection{Simulation Results}
 Our primary interest is selection (identification) of the relevant
 covariates $Z_1(\cdot), Z_2(\cdot), Z_3(\cdot)$ and estimating
 their effects $\beta_1(t), \beta_2(t), \beta_3(t)$ accurately. As the covariates are observed sparsely and with measurement error, we apply FPCA as discussed in Section \ref{sec:extension} with PVE= $99\%$ and obtain the denoised curves $\hat{Z}_{ij}(t)$ before applying our variable selection method. We apply the proposed variable selection method with and without the pre-whitening procedure mentioned in Section \ref{sec:prewhite}. Table \ref{tab1} and Table \ref{tab12} display the selection percentage of each variable for each of the three selection methods discussed in Section \ref{sec:method} and for the three sample sizes $n=100, 200, 400$, for the non pre-whitened and pre-whitened case respectively. We use the acronyms FLASSO (Functional LASSO), FSCAD (Functional SCAD) and FMCP (Functional MCP) respectively for the proposed variable selection methods for FLCM. We expect that the group LASSO selection method to have a higher false positive rate and use this as a benchmark for comparison.  It can be seen from Table \ref{tab1} and \ref{tab12} that all the three methods (group LASSO, group SCAD, group MCP) pick out the three true covariates $Z_1(\cdot), Z_2(\cdot), Z_3(\cdot)$;
$100\%$ of the time. The group LASSO method has a high false positive selection percentage as can be seen in both Table \ref{tab1} and Table \ref{tab12}, with selection accuracy improving with increasing sample size. The group SCAD and group MCP method, on the other hand, have a false selection percentage in the range of $0.2\%-1\%$ for non pre-whitened case and exactly $0\%$ for pre-whitened case. In other words, the group SCAD and group MCP method are able to identify the true model using the pre-whitening procedure. In scalar regression, SCAD and MCP are known to produce sparser solutions than LASSO due to its concave nature, and here also in the context of variable selection in functional linear concurrent regression, we observe these two methods (their group extension) outperforming LASSO. The average model sizes for each scenario are also given in Table \ref{tab1}, and the group SCAD and group MCP method produce smaller and closer values to the true model size 3 (exactly 3 with pre-whitening procedure in Table \ref{tab12}). These results also illustrate the benefit of pre-whitening and henceforth we have used pre-whitening as a preprocessing step to perform variable selection using the proposed methods. 

Next, as an assessment of the accuracy of the estimates $\hat{\beta}_k(t)$ ($k=1,2,3$), we plot the true regression curves overlaid by their Monte Carlo (MC) mean estimate from the three methods. MC point-wise confidence intervals ($95\%$) (corresponding to point-wise 2.5 and 97.5 percentiles of the estimated curves over 500 replicates) for each of the three curves are also displayed to

 { 
 \renewcommand{\arraystretch}{3}
\begin{landscape}
\begin{table}
\caption{\label{tab1} Comparison of selection percentages ($\%$) of different variables and average model size, without pre-whitening. }
\centering
\begin{adjustbox}{width=1.4\textwidth}
\fontsize{60}{22}\selectfont
\begin{tabular}{|c|c|c|c|c|c|c|c|c|c|c|c|c|c|c|c|c|c|c|c|c|c|c|}
\hline
Sample Size            & Method & Var1 & Var2 & Var3 & Var4 & Var5 & Var6 & Var7 & Var8 & Var9 & Var10 & Var11 & Var12 & Var13 & Var14 & Var15 & Var16 & Var17 & Var18 & Var19 & Var20 & \begin{tabular}[c]{@{}c@{}}Avg Model\\      Size\end{tabular} \\ \hline
\multirow{3}{*}{n=100} & FLASSO  & 100  & 100  & 100  & 16.4 & 18.4 & 16.6 & 10 & 14.4 & 15   & 15.6  & 17.2  & 15.2  & 13  & 14.2  & 17.8  & 16.4    & 15.4  & 16  & 14.4    & 13  & 5.59                                                         \\ \cline{2-23} 
                       & FSCAD   & 100  & 100  & 100  & 0.6  & 0.4  & 0.2  & 1.0  & 1.2  & 0.4  & 0.4   & 0.8   & 0.4   & 0.2   & 0.4   & .8     & 0.2   & 0.2   & 0.6   & 0     & 1     & 3.088                                                         \\ \cline{2-23} 
                       & FMCP    & 100  & 100  & 100  & 0.6  & 0.2  & 0.2  & 1.0  & 1.0  & 0.4  & 0.4   & 0.6   & 0.4   & 0.2   & 0.2   & 0.4   & 0.2   & 0.2   & 0.6   & 0     & 1     & 3.076                                                         \\ \hline
\multirow{3}{*}{n=200} & FLASSO  & 100  & 100  & 100  & 15.8 & 14.6   & 16.8 & 14   & 13.4 & 15.4 & 11.6  & 14.2  & 14.8    & 14.4    & 15  & 14.4  & 14  & 11.2  & 14.6  & 10.8  & 15  & 5.4                                                         \\ \cline{2-23} 
                       & FSCAD   & 100  & 100  & 100  & 0.2  & 0.6  & 1  & 0.6  & 0.4  & 0.8    & 0.8     & 1.2   & 0.2   & 0.2   & 0     & 0.4   & 0.2   & 0.4   & 0.8   & 0.4   & 1     & 3.092                                                         \\ \cline{2-23} 
                       & FMCP    & 100  & 100  & 100  & 0.2  & 0.6  & 0.8  & 0.4  & 0.4  & 0.8    & 0.6   & 1   & 0.2   & 0.2   & 0     & 0.4   & 0.2   & 0.4   & 0.8   & 0.2   & 0.8     & 3.08                                                         \\ \hline
\multirow{3}{*}{n=400} & FLASSO  & 100  & 100  & 100  & 13.8 & 13.4   & 14.8 & 11.2 & 12.6 & 12.8 & 10.8    & 13.4  & 11  & 12.4    & 12.4  & 15.2  & 11.2  & 12.8  & 13.2  & 12  & 13.6  & 5.176                                                         \\ \cline{2-23} 
                       & FSCAD   & 100  & 100  & 100  & 0.4  & 0  & 0.2  & 0.4  & 0.4  & 0    & 0   & 0   & 0     & 0.4   & 0     & 0.2   & 0.2   & 0.2   & 0.2   & 0     & 0   & 3.026                                                          \\ \cline{2-23} 
                       & FMCP    & 100  & 100  & 100  & 0.4  & 0  & 0.2  & 0.4  & 0.2  & 0    & 0   & 0   & 0     & 0.4   & 0     & 0.2   & 0.2   & 0.2   & 0.2   & 0     & 0   & 3.024                                                          \\ \hline
\end{tabular}
\end{adjustbox}
\end{table}
\end{landscape}
}
\clearpage 
\newpage

\begin{figure}[H]
\begin{center}
    \begin{tabular}{ll}
        \scalebox{0.58}{\includegraphics{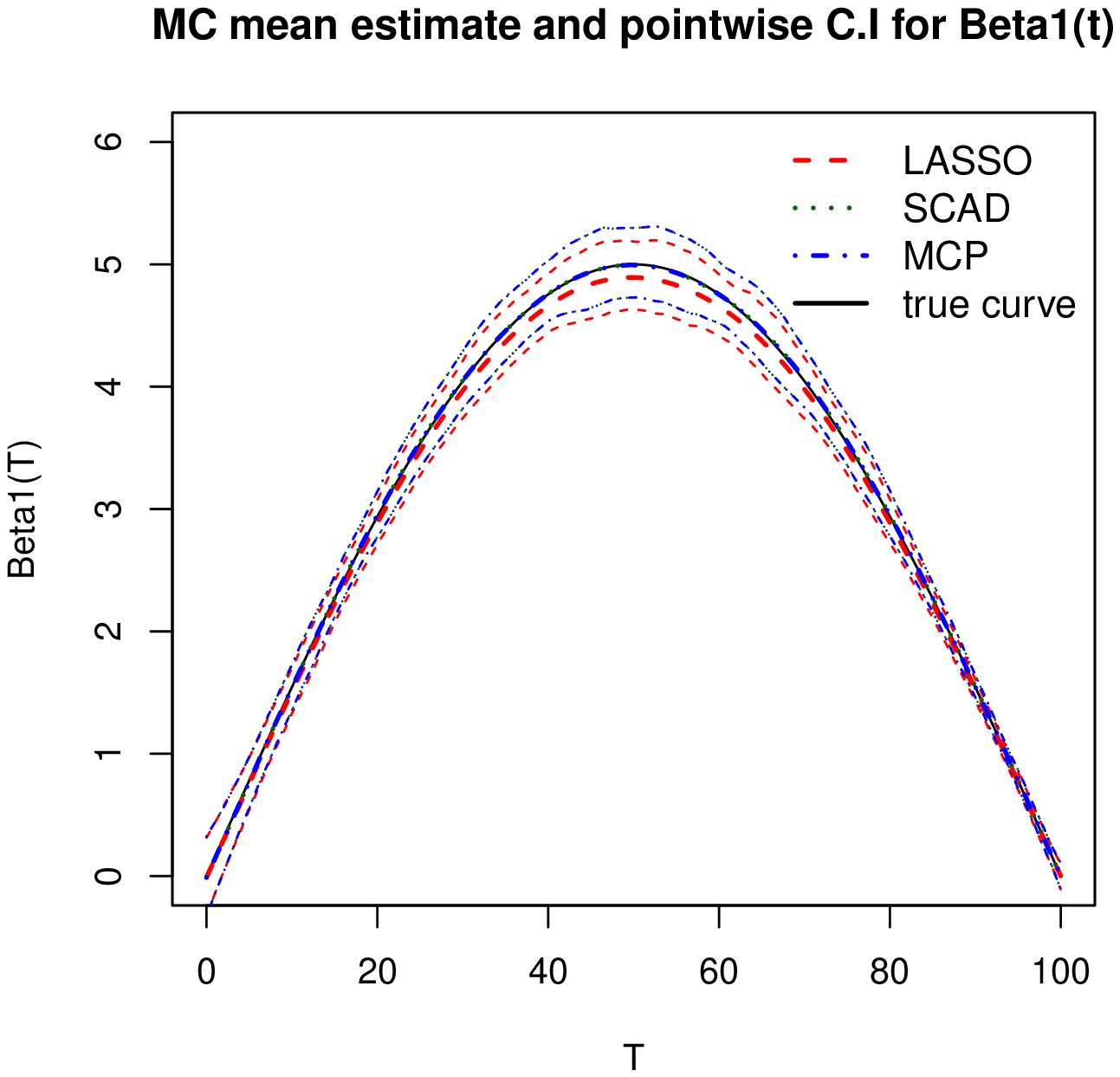}} &
 \scalebox{0.58}{\includegraphics{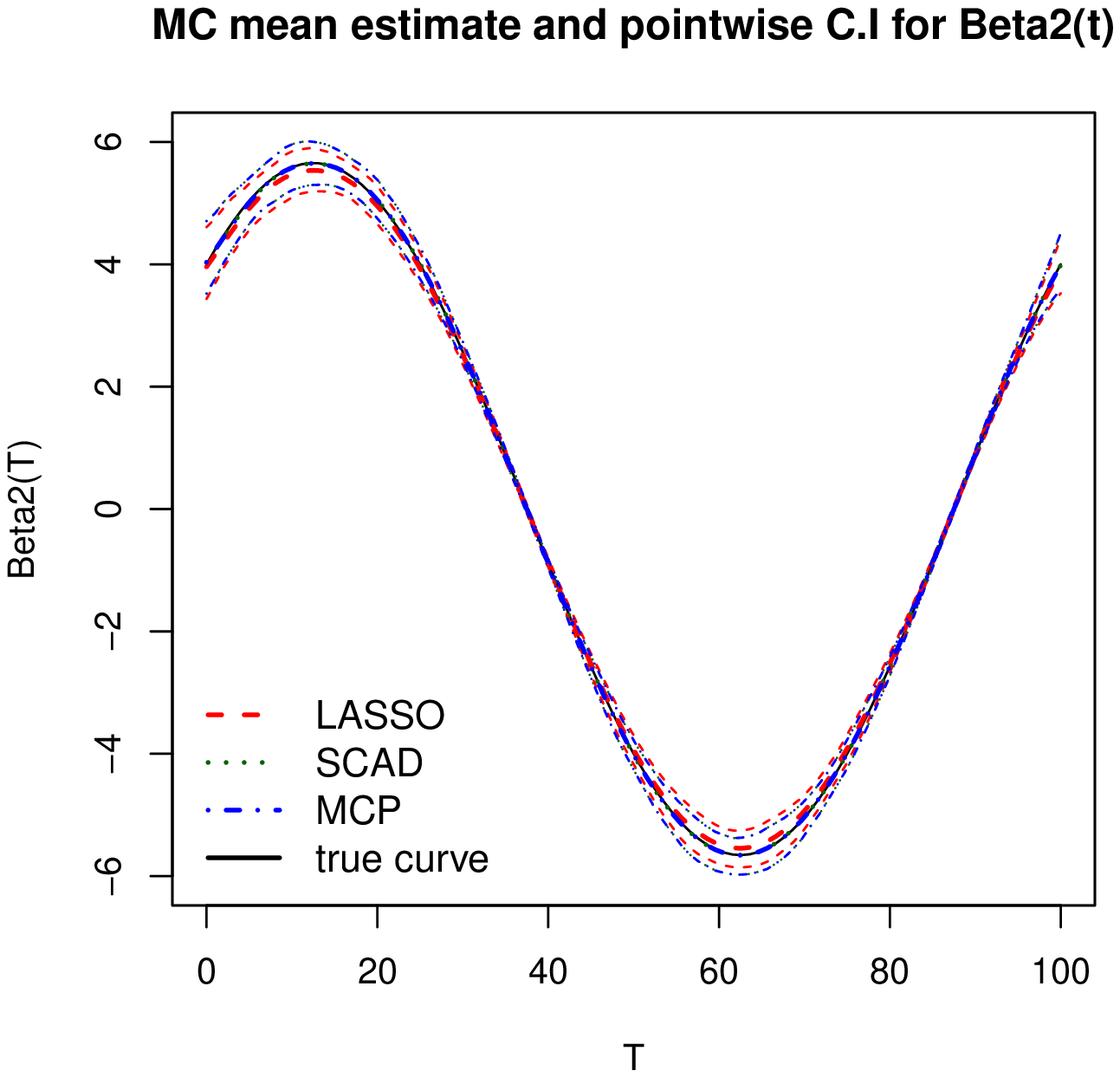}}\\
\scalebox{0.58}{\includegraphics{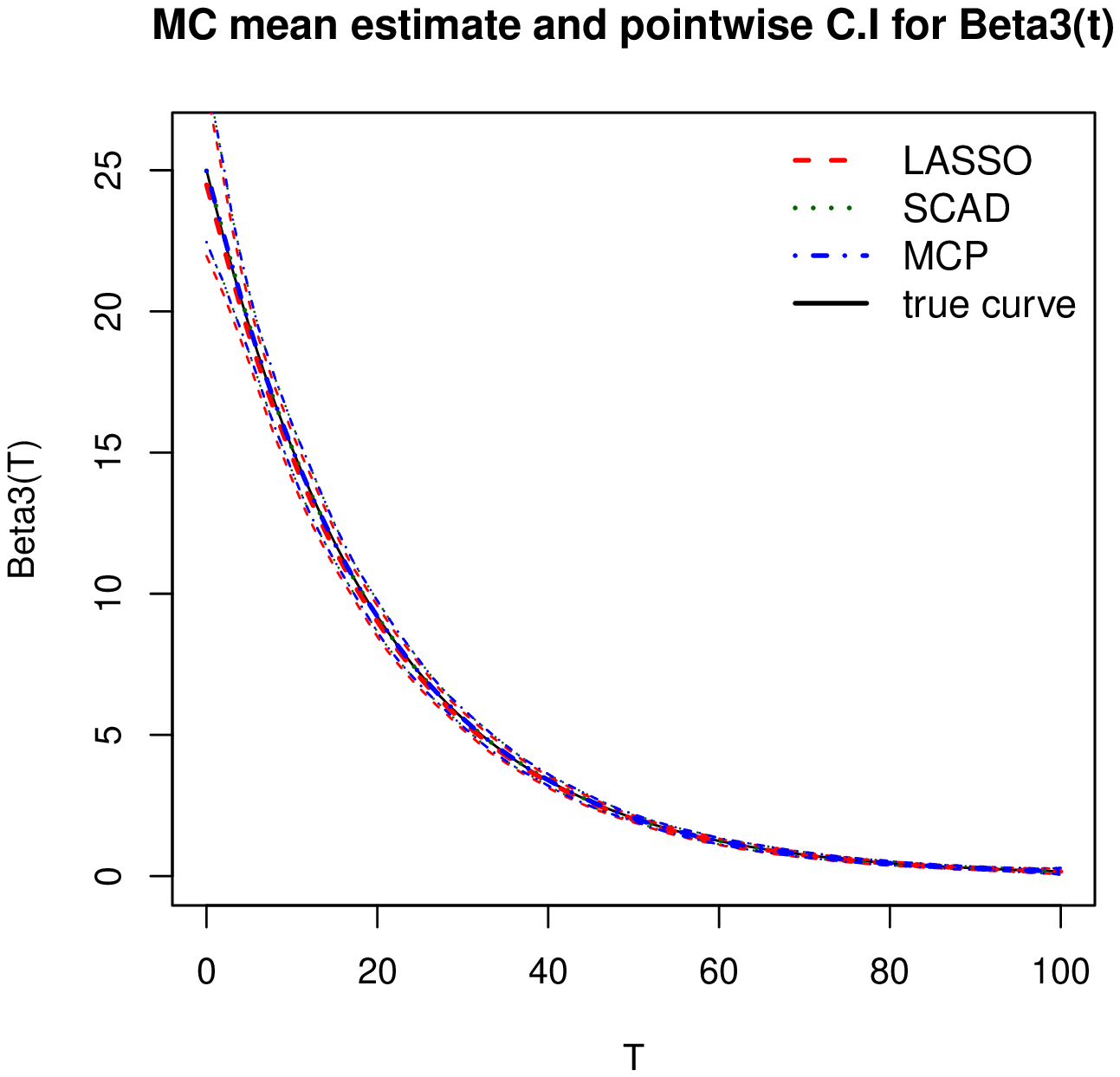}} &
\\
    \end{tabular}
\end{center}
\caption{MC estimates and pointwise confidence intervals of the coefficient functions (n=200).}
\label{fig:fig1}
\end{figure}

{ 
 \renewcommand{\arraystretch}{3}
\begin{landscape}
\begin{table}
\caption{\label{tab12} Comparison of selection percentages ($\%$) of different variables and average model size, with pre-whitening. }
\centering
\begin{adjustbox}{width=1.4\textwidth}
\fontsize{60}{22}\selectfont
\begin{tabular}{|c|c|c|c|c|c|c|c|c|c|c|c|c|c|c|c|c|c|c|c|c|c|c|}
\hline
Sample Size            & Method & Var1 & Var2 & Var3 & Var4 & Var5 & Var6 & Var7 & Var8 & Var9 & Var10 & Var11 & Var12 & Var13 & Var14 & Var15 & Var16 & Var17 & Var18 & Var19 & Var20 & \begin{tabular}[c]{@{}c@{}}Avg Model\\      Size\end{tabular} \\ \hline
\multirow{3}{*}{n=100} & FLASSO  & 100  & 100  & 100  & 6.2 & 7.8 & 7.6 & 6 & 7.8 & 6.4   & 6.6  & 7.4  & 5.8  & 5.4  & 6.4  & 6.4  & 7.4    & 4.8  & 6  & 6    & 6.2  & 4.112                                                         \\ \cline{2-23} 
                       & FSCAD   & 100  & 100  & 100  & 0  & 0  & 0  & 0  & 0  & 0  & 0   & 0   & 0   & 0   & 0   & 0    & 0   & 0   & 0   & 0     & 0     & 3                                                         \\ \cline{2-23} 
                       & FMCP    & 100  & 100  & 100  & 0  & 0  & 0  & 0  & 0  & 0  & 0   & 0   & 0   & 0   & 0   & 0    & 0   & 0   & 0   & 0     & 0     & 3                                                       \\ \hline
\multirow{3}{*}{n=200} & FLASSO  & 100  & 100  & 100  & 5.8 & 3 & 6   & 4.4 & 5.2   & 4.8 & 5.6 & 4.2  & 7.6  & 4.4    & 4.8    & 3.2  & 3.4  & 2.8  & 5.6  & 4.4  & 16  & 3.932                                                         \\ \cline{2-23} 
                       & FSCAD   & 100  & 100  & 100  & 0  & 0  & 0  & 0  & 0  & 0  & 0   & 0   & 0   & 0   & 0   & 0    & 0   & 0   & 0   & 0     & 0     & 3                                                      \\ \cline{2-23} 
                       & FMCP   & 100  & 100  & 100  & 0  & 0  & 0  & 0  & 0  & 0  & 0   & 0   & 0   & 0   & 0   & 0    & 0   & 0   & 0   & 0     & 0     & 3                                                     \\ \hline
\multirow{3}{*}{n=400} & FLASSO  & 100  & 100  & 100  & 4.2 & 2.6   & 4.6 & 3.8 & 3.6 & 3 & 2.6    & 3.4  & 5  & 5.2    & 5.4  & 3.6  & 3.4  & 3.4  & 4.6  & 4.8  & 29  & 3.922                                                         \\ \cline{2-23} 
                       & FSCAD   & 100  & 100  & 100  & 0  & 0  & 0  & 0  & 0  & 0  & 0   & 0   & 0   & 0   & 0   & 0    & 0   & 0   & 0   & 0     & 0     & 3                                                           \\ \cline{2-23} 
                       & FMCP    & 100  & 100  & 100  & 0  & 0  & 0  & 0  & 0  & 0  & 0   & 0   & 0   & 0   & 0   & 0    & 0   & 0   & 0   & 0     & 0     & 3                                                           \\ \hline
\end{tabular}
\end{adjustbox}
\end{table}
\end{landscape}
}
\clearpage 
\newpage
\hspace*{ -6 mm}
asses variability of the estimates. Figure \ref{fig:fig1} displays this plot for $n=200$, the plots for $n=100,n=400$ are similar with more accuracy and less variability for larger sample sizes. The group LASSO estimates (dashed line) have a larger bias which is again expected, as LASSO is known to have a relatively high bias when magnitude of the regression coefficient is large. The group SCAD (dotted line) and group MCP (dashed-dotted line) estimates have almost identical accuracy and variability as seen from Figure \ref{fig:fig1}; they have superimposed on each other and on the true curves represented by solid lines. 
\clearpage
\newpage

To further evaluate the performance of the estimates we calculate the absolute bias and the MC mean square error of the estimates averaged across $100$ equally spaced grid points in $[0,100]$, for all the selection methods and the three sample sizes. This is displayed in Table \ref{tab2}. We again observe group SCAD and group MCP method outperforming the group LASSO method, in terms of both absolute bias and mean square error, the performance of the estimators improving with increasing sample size. We compared these mean square errors of the estimates arising from pre-whitening procedure with the same from non pre-whitening procedure and found these only to be marginally higher, which is expected due to the uncertainty associated with estimating the covariance matrix. The mean square errors appear to be converging to zero across all the three methods with increase in sample size indicating consistency of the estimators. The simulation results illustrate superior performance of the proposed group SCAD (FSACD) or group MCP (FMCP) based selection method in the context of functional linear concurrent model and are the recommended methods of this article.

\begin{table}
\caption{\label{tab2} Comparison of MC absolute bias and mean square error.}
\centering
\large
\begin{tabular}{|c|c|c|c|c|c|c|c|}
\hline
\multirow{2}{*}{\begin{tabular}[c]{@{}c@{}}Sample\\ Size\end{tabular}} & \multirow{2}{*}{Method} & \multicolumn{2}{c|}{$\hat{\beta}_1(t)$} & \multicolumn{2}{c|}{$\hat{\beta}_2(t)$} & \multicolumn{2}{c|}{$\hat{\beta}_3(t)$} \\ \cline{3-8} 
                                                                       &                         & Bias               & MSE                & Bias               & MSE                & Bias               & MSE                \\ \hline
\multirow{3}{*}{n=100}                                                 & FLASSO                   & 0.083              & 0.033              & 0.092              & 0.048              & 0.112              & 0.191              \\ \cline{2-8} 
                                                                       & FSCAD                    & 0.011              & 0.025              & 0.015              & 0.038              & 0.022              & 0.165              \\ \cline{2-8} 
                                                                       & FMCP                     & 0.011              & 0.025              & 0.015              & 0.038              & 0.022              & 0.165              \\ \hline
\multirow{3}{*}{n=200}                                                 & FLASSO                   & 0.061              & 0.017              & 0.069              & 0.024              & 0.092              & 0.109              \\ \cline{2-8} 
                                                                       & FSCAD                    & 0.007              & 0.013              & 0.008              & 0.019              & 0.010              & 0.091              \\ \cline{2-8} 
                                                                       & FMCP                     & 0.007              & 0.013              & 0.008              & 0.019              & 0.010              & 0.091              \\ \hline
\multirow{3}{*}{n=400}                                                 & FLASSO                   & 0.047              & 0.009              & 0.051              & 0.013              & 0.070              & 0.063              \\ \cline{2-8} 
                                                                       & FSCAD                    & 0.004              & 0.007              & 0.004              & 0.010              & 0.010              & 0.050              \\ \cline{2-8} 
                                                                       & FMCP                     & 0.004              & 0.007              & 0.004              & 0.010              & 0.010              & 0.050              \\ \hline
\end{tabular}
\end{table}
\section{Real Data Applications}
In this section, we demonstrate application of our variable selection method in selection of influential time-varying predictors in two real data studies. For performing variable selection, we use only the FSCAD and FMCP method along with the initial pre-whitening procedure, as the group LASSO method yields a significantly higher false positive rate which is illustrated by our simulations. We first consider a small dietary calcium absorption dataset (three time-varying covariates) with added pseudo covariates as an illustration of our method.
Addition of pseudo covariates is a popular way \citep{wang2008variable, wu2007controlling, miller2002subset} of assessing false selection rate in real datasets. Pseudo variables can, therefore, be used effectively for tuning variable selection procedures.
We show that our proposed method is able to select the relevant predictors and discard the pseudo variables successfully. Finally, we apply our variable selection method to the fisheries dataset to find out relevant socio-economic drivers influencing fisheries footprint of nations over time.
\label{sec:real_dat}
\subsection{Study of Dietary Calcium Absorption}
\begin{figure}[ht]
\centering
\includegraphics[width=.9\linewidth , height=.8\linewidth]{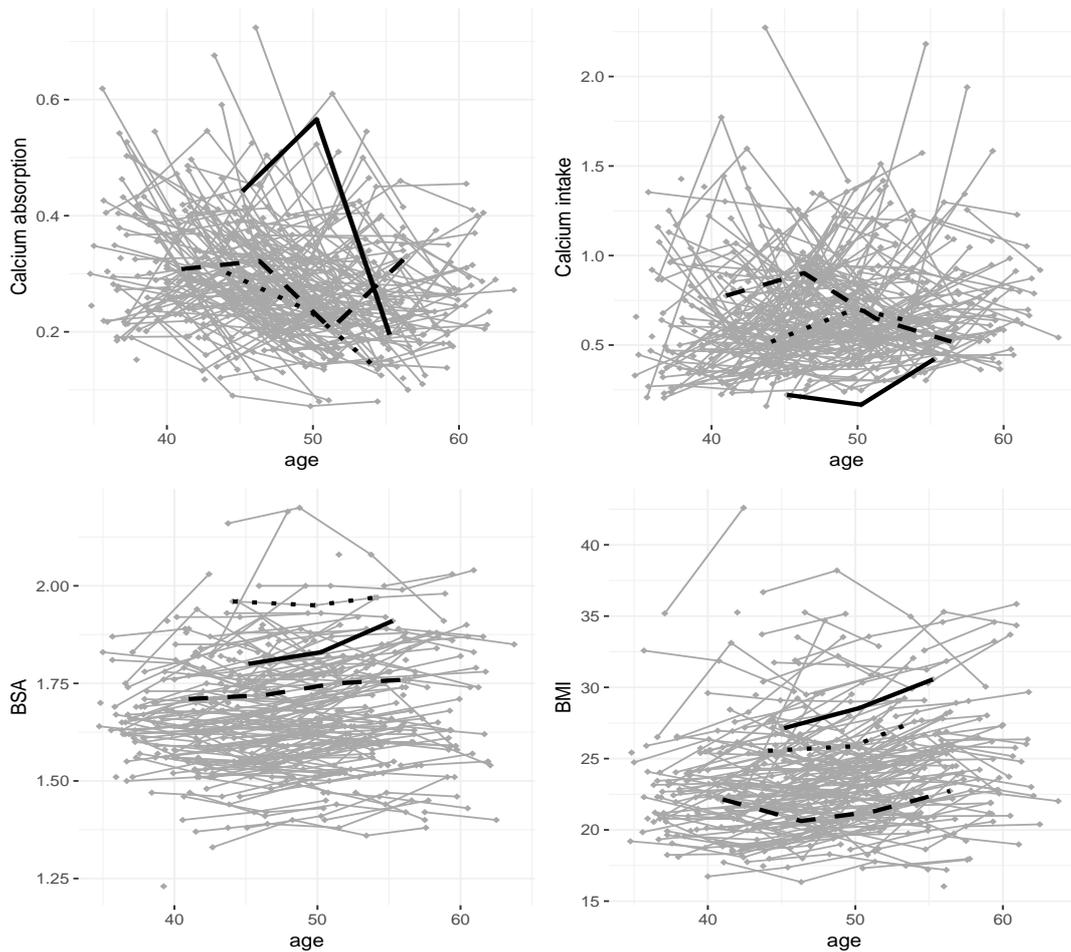}
\caption{Calcium absorption and covariate profiles of patients along their ages.}
\label{fig:fig2}
\end{figure}

 We consider the study of dietary calcium absorption in \cite{davis2002statistical}. In this study, the subjects are a group of 188 patients. We have data on calcium absorption ($Y(t)$), dietary calcium intake ($Z_1(t)$), BMI ($Z_2(t)$) and BSA (Body surface area) ($Z_3(t)$) of these patients, at irregular time points between 35 and 64 years of their ages. At the beginning of the study patients aged between 35 and 45 years and subsequent observations were taken approximately every 5 years. The number of repeated measurements for each patient varies from 1 to 4. Figure \ref{fig:fig2} displays the individual curves of patients' calcium absorption, calcium intake, BSA, BMI along their ages.
 
 We are primarily interested in finding out which covariates influence calcium absorption profile of the patients. \cite{kim2016general} also investigated the effect of calcium intake on calcium absorption using an additive nonlinear functional concurrent model, and found the effect to be more or less linear while comparing to a functional linear concurrent model. So we use functional linear concurrent regression to model the dependence of calcium absorption on calcium intake, BSA and BMI. As data is observed very sparsely and the original covariates might be observed with measurement error, we apply FPCA methods (PVE = $95\%$) as discussed in Section \ref{sec:extension} and get the denoised trajectories $\hat{Z}_{j}(t)$ for $j=1,2,3$. We expect that, calcium intake among the three covariates will be associated with calcium absorption. We add 15 pseudo covariates by simulating from the following functional model to illustrate the selection performance and false positive rate of our variable selection method. We generate $Z_{ij}(\cdot)\stackrel{iid}{\sim} Z_j(\cdot)$ where $Z_j(t)$ ($j=4,5,\ldots,18$) are given by $Z_j(t)=a_j\hspace{1mm}\sqrt[]{2}sin(\pi (j-3) t/200)+ b_j\hspace{1mm}\sqrt[]{2}cos(\pi (j-3) t/200)$, where $a_j\sim \mathcal{N}(0,(2)^2)$, $b_j\sim \mathcal{N}(0,(2)^2)$.
 So in total, we have 18 covariates, where the first 3 are the denoised original covariates and rest are simulated predictors. Then we apply our variable selection method to $Y(t)$ and $\hat{Z}_{1}(t),\hat{Z}_{2}(t),\hat{Z}_{3}(t),Z_4(t),Z_5(t),\ldots,Z_{18}(t)$. We repeat this a large number of times and observe which variables are being selected in each iteration. We expect our variable selection method to pick out the truly influential predictors and ignore the randomly generated functional covariates the majority of the time. To illustrate the benefit of using our proposed variable selection method in functional regression model for this particular data we compare its performance to a backward selection method which uses model selection criterion like BIC or Mallows' Cp, under a linear model approach (using an independent working correlation structure), and to a penalized generalized estimating equations (PGEE) procedure \citep{wang2012penalized} which was developed to analyze longitudinal data with a large
number of covariates. We use the `PGEE' package in R \citep{inan2017pgee} for implementing the penalized generalized estimating
equations procedure under two different working correlation structure (independent and AR (1)). 

Table \ref{tab3} illustrates the selection percentage of each of the variables under different methods. We notice that both the proposed FSCAD and FMCP method identify calcium intake ($Z_1(t)$) as a significant predictor $100\% $ of the time. All other variables including all the pseudo covariates are ignored in $100\%$ of the iterations. On the other hand, BIC, Cp, and PGEE exhibit a high false selection percentage for the pseudo variables. The case of selection of BSA or BMI appears to be over selection as their individual effects were not found to be statistically significant. This demonstrates when the underlying model is functional, the use of naive variable selection methods using scalar regression techniques can lead to wrong inference as they don't account for functional nature of the data.

\begin{table}
\caption{\label{tab3} Selection Percentages ($\%$) of variables in Calcium absorption Study.}
\centering
\hspace*{- 1 mm}
\large
\begin{tabular}{|c|c|c|c|c|}
\hline
Method & Var 1(Calcium Intake) & Var 2 (BSA) & Var 3(BMI) & Max Var (4-18) \\ \hline
FSCAD    & 100                   & 0           & 0          & 0   \\ \hline
FMCP     & 100                   & 0           & 0          & 0     \\ \hline
BIC (LM)    & 100                   & 100           & 0          & 9     \\ \hline
Cp (LM)    & 100                   & 100           & 2          & 33     \\ \hline
PGEE (ind)    & 100                   & 0           & 100          & 31     \\ \hline
 PGEE (AR-1)    & 100                   & 0           & 100         & 31     \\ \hline
\end{tabular}
\end{table}

\begin{figure}[ht]
\begin{center}
\begin{tabular}{ll}
 \scalebox{0.4}{\includegraphics{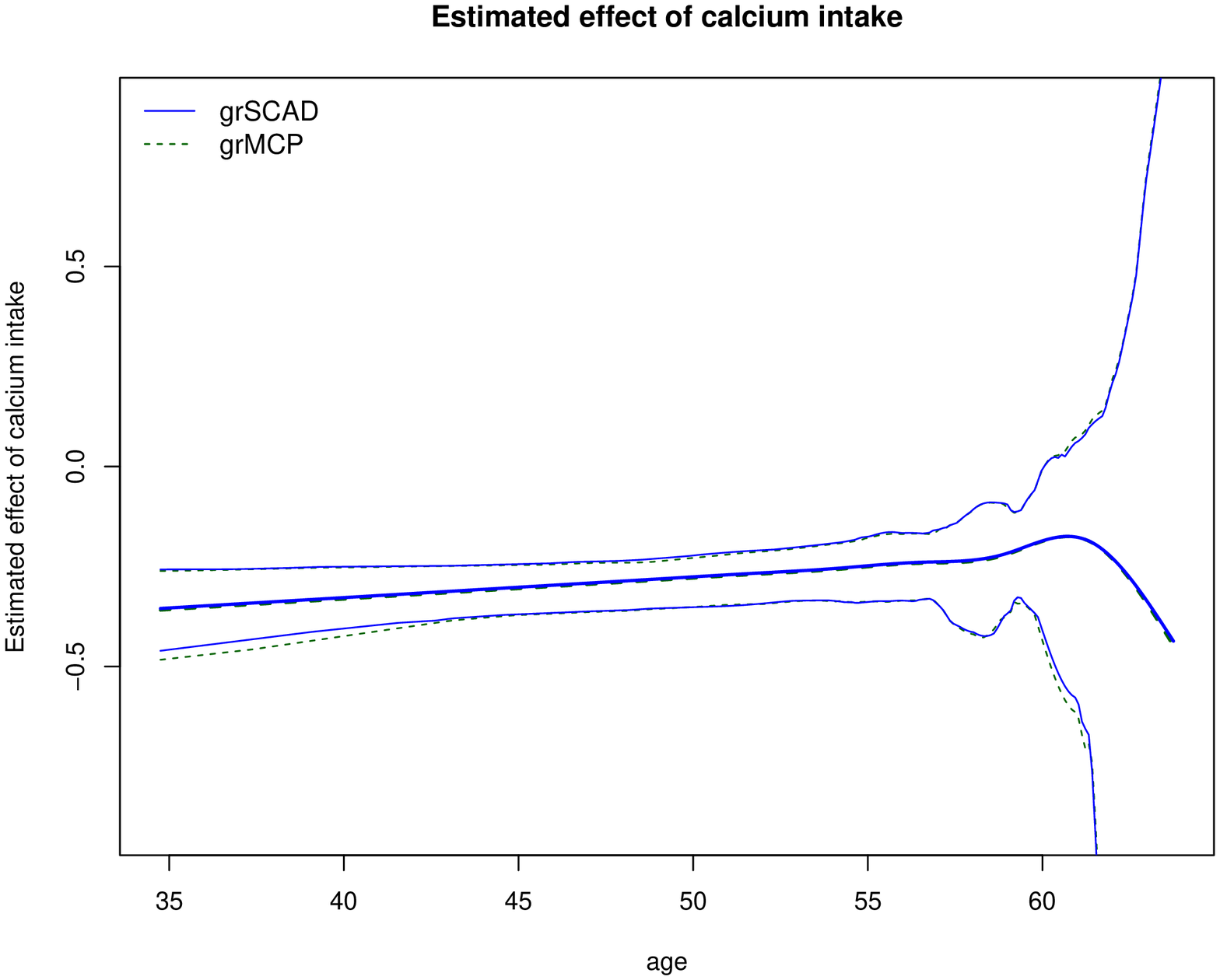}} &
 \scalebox{0.4}{\includegraphics{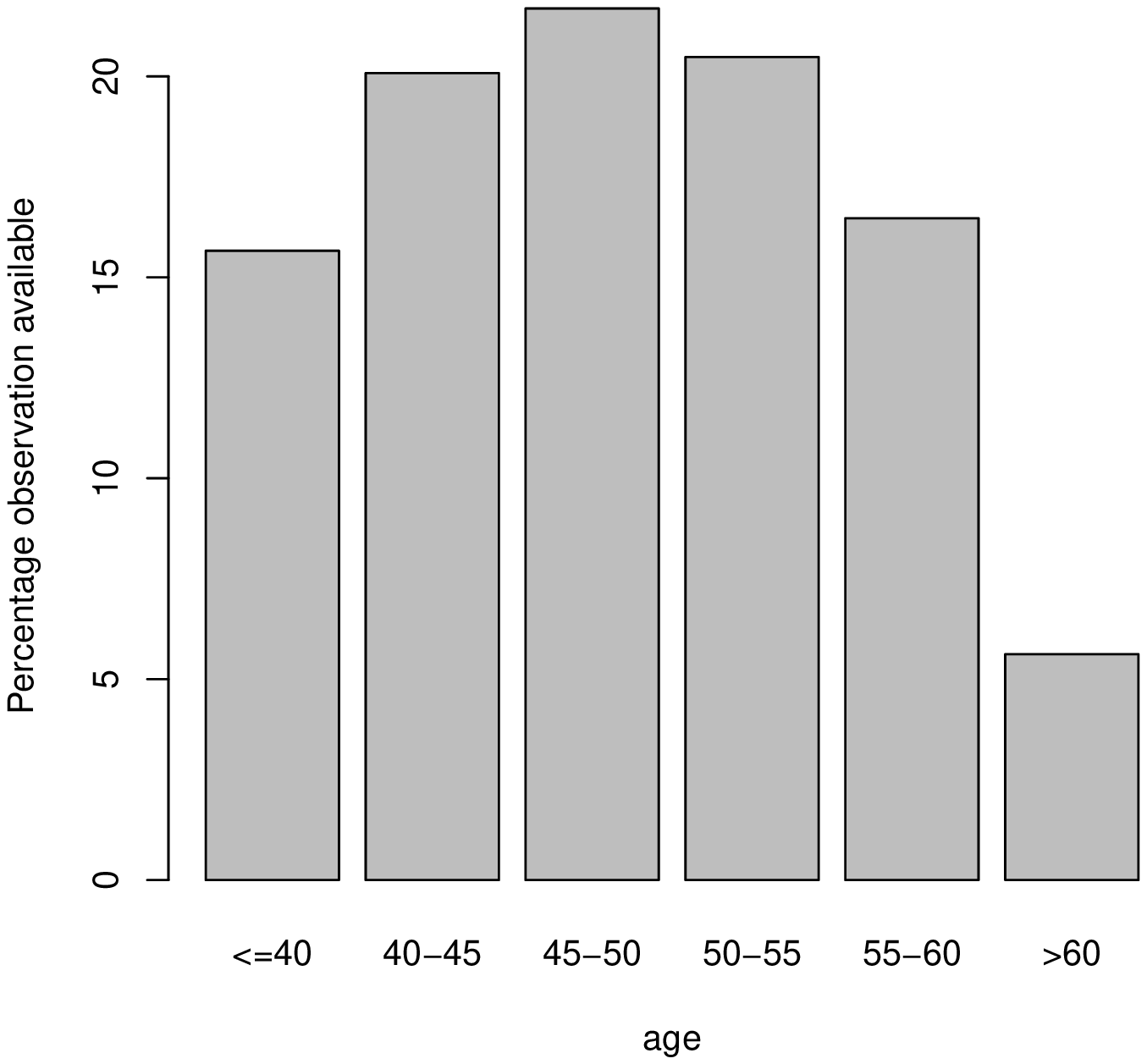}}\\
\end{tabular}
\end{center}
\caption{Bootstrap estimate and pointwise confidence interval of the the effect of calcium intake (left panel) and percentage observation available in different age groups (right panel).}
\label{fig:fig3}
\end{figure}

As calcium intake is the only significant variable selected by both the proposed methods we want to estimate its effect and also get a measure of uncertainty of our estimate. For this purpose, we use a subject-level bootstrap on our original data (no pseudo covariates added) while performing variable selection to come up with an estimated regression curve $\hat{\beta}_1(t)$ and a pointwise confidence interval for the effect of calcium intake. This is displayed in Figure \ref{fig:fig3}. We notice as calcium intake increase calcium absorption should decrease particularly until age $60$ years, as $\hat{\beta}_1(t)<0$ up to this age and the confidence interval strictly lies below zero, which might be due to dietary calcium saturation or due to interaction with some other elements in the body; although the overall magnitude of the effect seems to decrease with age. Above age $60$, the estimate appears to have high variability associated with it, which is primarily because we have very few observations ($5.62\%$) above this mark (illustrated in Figure \ref{fig:fig3}). The uncertainty in estimating $\hat{\#\Sigma}$ using FPCA and the uncertainty due to bootstrap is reflected in its variability. Hence, some care should be taken in interpretation of the estimated regression curve beyond 60 years because of such high uncertainty.
\subsection{Study of Fisheries Footprint}
\label{ffp}
Production of fisheries is a source of protein as well as an economic livelihood across the world. Along with the increasing global population, the importance of fish production and consumption has steadily increased through the modern era. Fisheries Footprint is defined as the Global Footprint Network's measure of total marine area required to sustain consumption levels of aquatic production of fish, crustacean (e.g., shrimp), shellfish, and seaweed from captures and aquaculture; so the fisheries footprint basically represents the coastal and marine area required to sustain the amount of seafood products a nation consumes. As pointed out by \cite{longo2016ocean}, the interaction between marine and social systems calls for further sociological analysis.

Over the last two decades, social scientists have accomplished much in advancing scholarly knowledge on the social drivers of ecological impact at a macro-scale. Such work is essential, as ecological problems are becoming increasingly interlinked and severe at a global or planetary scale \citep{steffen2011anthropocene}. Over time, for example, economic development, population structuring (e.g., urbanization or age structure), trade relations, and technological change are shown to affect measures of environmental impact across nations over time \citep{doi:10.1177/0020715219869976, jorgenson2010assessing, york2003footprints}. This body of literature centers on the ecological affects of globalization and modernization, under the socio-structural parameters of a capitalist economy. There is still much debate over the impacts of industrial and agricultural modernization on ecosystems. For example, development and resource economics literature \citep{world2007} advocate for the utilization of innovation and techno-improvements to improve marine system sustainability and food security \citep{valderrama2010market}, while, on the other hand, environmental sociologists demonstrate that such innovation, chiefly aquaculture, does not displace the deleterious ecological impacts of capture fisheries \citep{longo2019aquaculture}. Nevertheless, despite such progress, fisheries footprint remains an understudied metric, and its drivers are less understood in social research \citep{clark2018socio,jorgenson2005unpacking}.

\begin{figure}
\centering
\includegraphics[width=.7\linewidth , height=.7\linewidth]{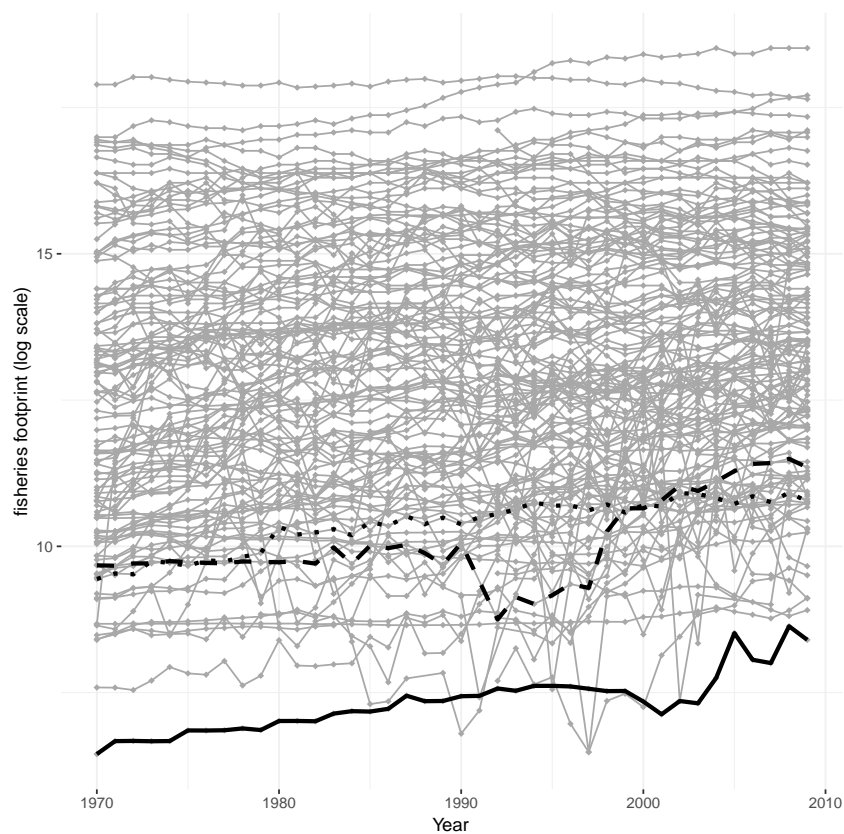}
\caption{Fisheries footprint of the nations over Year 1970-2009.}
\label{fig:fish1}
\end{figure}

\begin{figure}[ht]
\centering
\includegraphics[width=.7\linewidth , height=.7\linewidth]{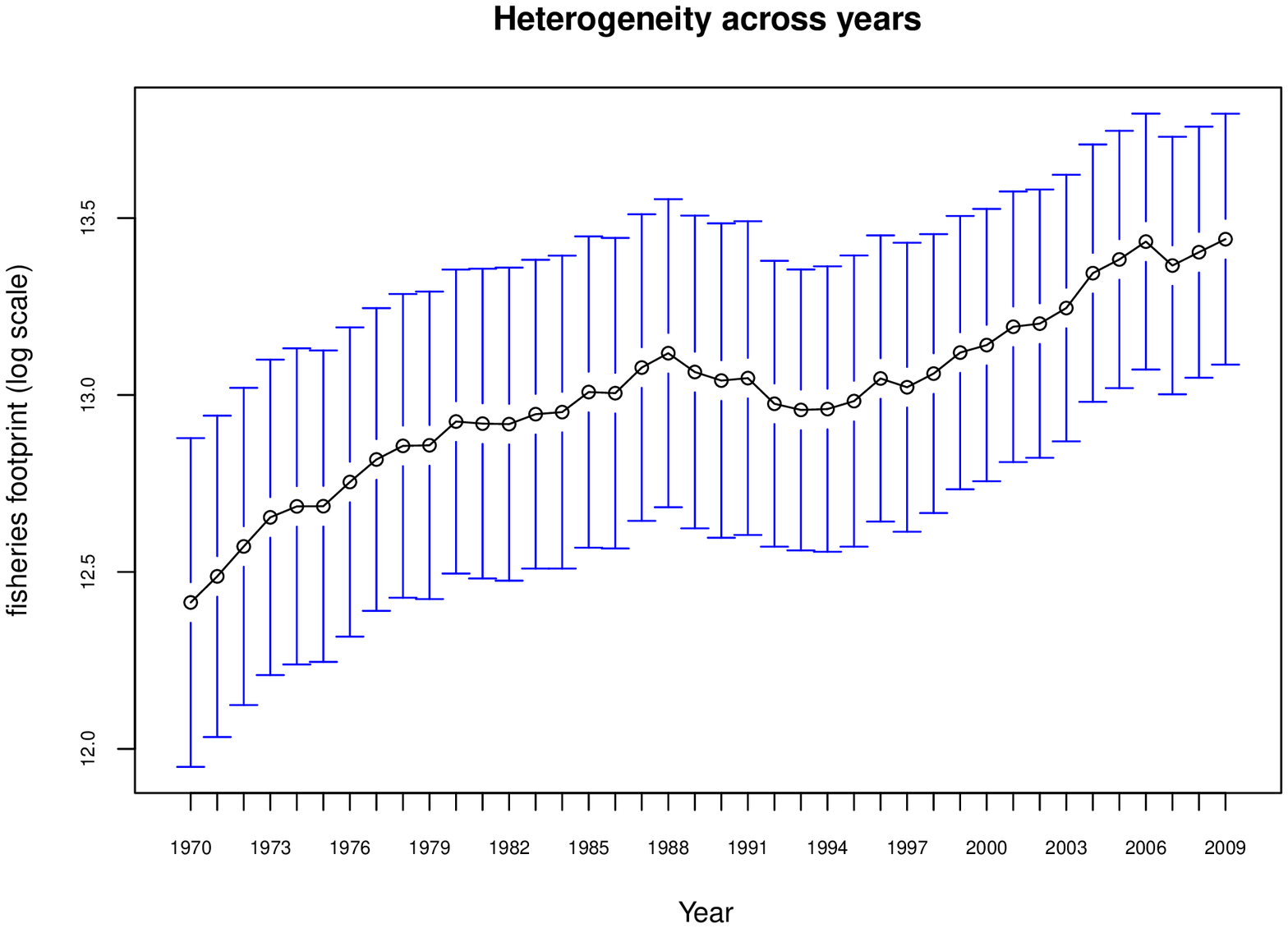}
\caption{Mean fisheries footprint along with their 95\% confidence interval. }
\label{fig:fish2}
\end{figure}

The goal of this study, therefore, is to identify the relevant socio-economic drivers such as levels of economic development, population size, and transformations in food-system dynamics that influence fisheries footprint of nations over time and also to capture their time-varying effects. Data for this study is collected from the World Bank, Fish StatJ of UN FAO, and Ecological Footprint Network for years between 1970-2009, across 136 nations. The main dependent variable of interest in this study is fisheries footprint. Figure \ref{fig:fish1} displays the fisheries footprint of the nations over the study years in log scale. Fisheries footprint of three representative nations are plotted using solid, dashed and dotted lines.
To capture the trend over the years, we plot the mean fisheries footprint of the nations along with their pointwise $95\%$ confidence interval. This is displayed in Figure \ref{fig:fish2}. We notice an overall upward trend as well as heterogeneity across years.

There are 20 independent time-varying covariates in the study, broadly covering various sectors of population dynamics (e.g. population density, urban population, total population, working age population percentage etc), agriculture (e.g. tractor, agriculture value added etc), food consumption and other fisheries variables (e.g. meat consumption, aquaculture production tons etc), international relations (e.g. food export as percentage of merchandise, FDI inflow etc), and economy (e.g. GDP per capita at constant U.S. dollar , trade percentage of GDP etc). The full list of the variables are given in Table \ref{tab4}. 
\begin{table}
\caption{\label{tab4} List of covariates in the Fisheries Footprint Study.}
\centering
\begin{tabular}{ll}
\hline
\multicolumn{2}{|l|}{\hspace{9mm} Predictor Variables in the Fisheries Footprint study}                                   \\ \hline
\multicolumn{1}{|l|}{agriculture value added}         & \multicolumn{1}{l|}{aquaculture production tons}     \\ \hline
\multicolumn{1}{|l|}{arable land hectares}            & \multicolumn{1}{l|}{arable land pct}                 \\ \hline
\multicolumn{1}{|l|}{exportsofgoodsandservicesofgdpn} & \multicolumn{1}{l|}{fao livestock}                   \\ \hline
\multicolumn{1}{|l|}{FDI inflow}                      & \multicolumn{1}{l|}{foodexportsofmerchandiseexprtst} \\ \hline
\multicolumn{1}{|l|}{foodimportsofmerchandiseimprtst} & \multicolumn{1}{l|}{gdp pc 2010}                     \\ \hline
\multicolumn{1}{|l|}{manufacturing value pctGDP}      & \multicolumn{1}{l|}{meat consumption FAO}            \\ \hline
\multicolumn{1}{|l|}{population 15\_64 pct}           & \multicolumn{1}{l|}{population density}              \\ \hline
\multicolumn{1}{|l|}{populationtotalsppoptotl}        & \multicolumn{1}{l|}{services value growth pct}       \\ \hline
\multicolumn{1}{|l|}{tractors}                        & \multicolumn{1}{l|}{trade pct GDP}                   \\ \hline
\multicolumn{1}{|l|}{urban pop}                       & \multicolumn{1}{l|}{urban pop pct}                   \\ \hline
 \end{tabular}
\end{table}
The predictors here are also time-varying and can be expected to have dynamic effects on fisheries footprint. Generally panel data methods like fixed effects or random effects modelling is used \citep{torres2007panel,doi:10.1177/0020715219869976} for analyzing such data where the effect the covariates are taken to be constant, since we are interested
in dynamic effects of the covariates, the FLCM can be seen as a generalization of this approach with time-varying effects of predictors. Therefore we use a functional linear concurrent regression model (1) discussed in this article to model the dynamic effects of the socio-economic predictors on fisheries footprint. The predictors in their original scale are also very large in magnitude, therefore converted into log scale; the covariates observed as percentages are used without any conversion. Before applying our variable selection method all the covariates are preprocessed using FPCA methods (PVE = $95\%$) as discussed in Section \ref{sec:extension}.

\begin{figure}[ht]
\begin{center}
\begin{tabular}{ll}
 \scalebox{0.5}{\includegraphics{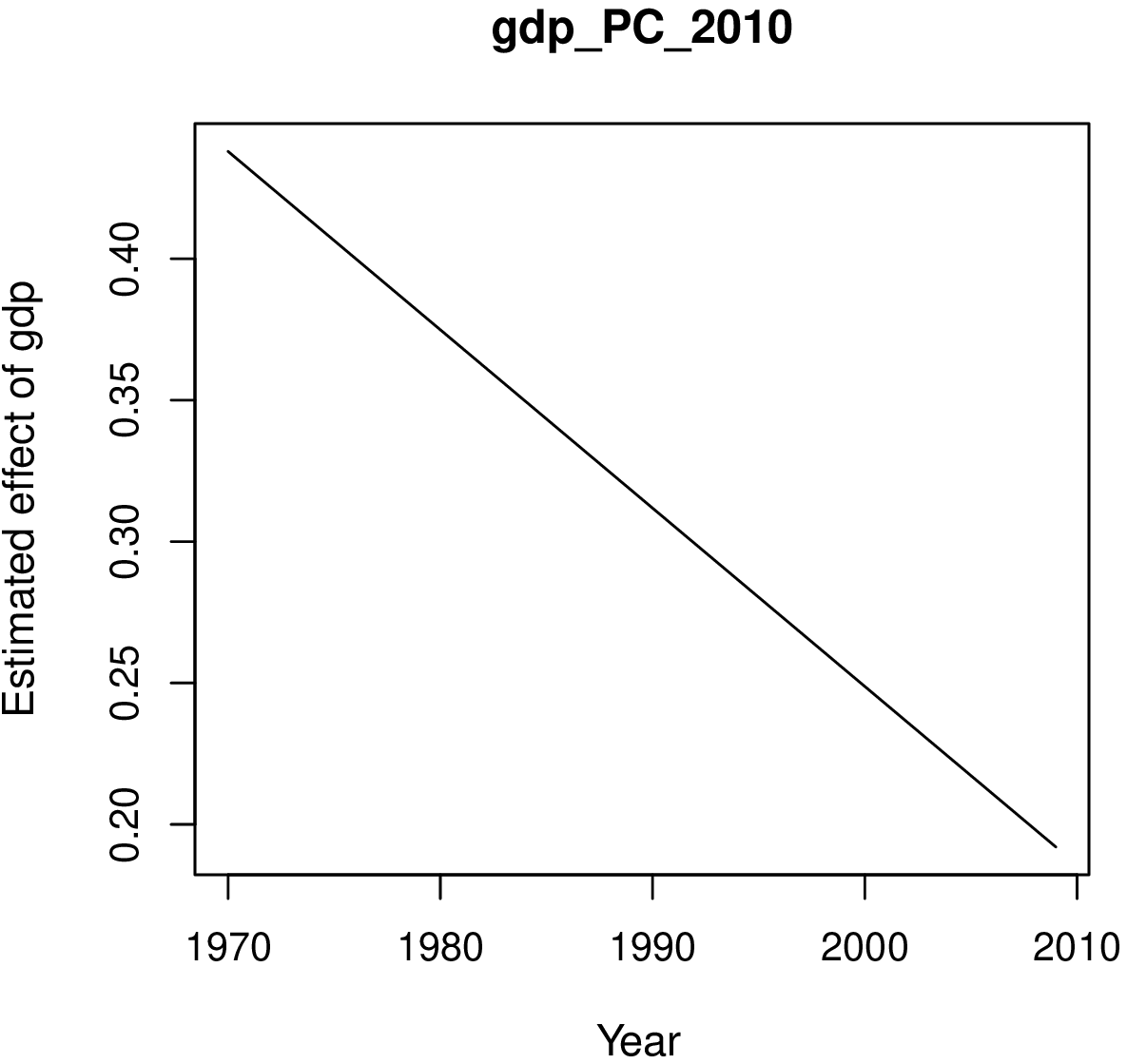}} &
 \scalebox{0.5}{\includegraphics{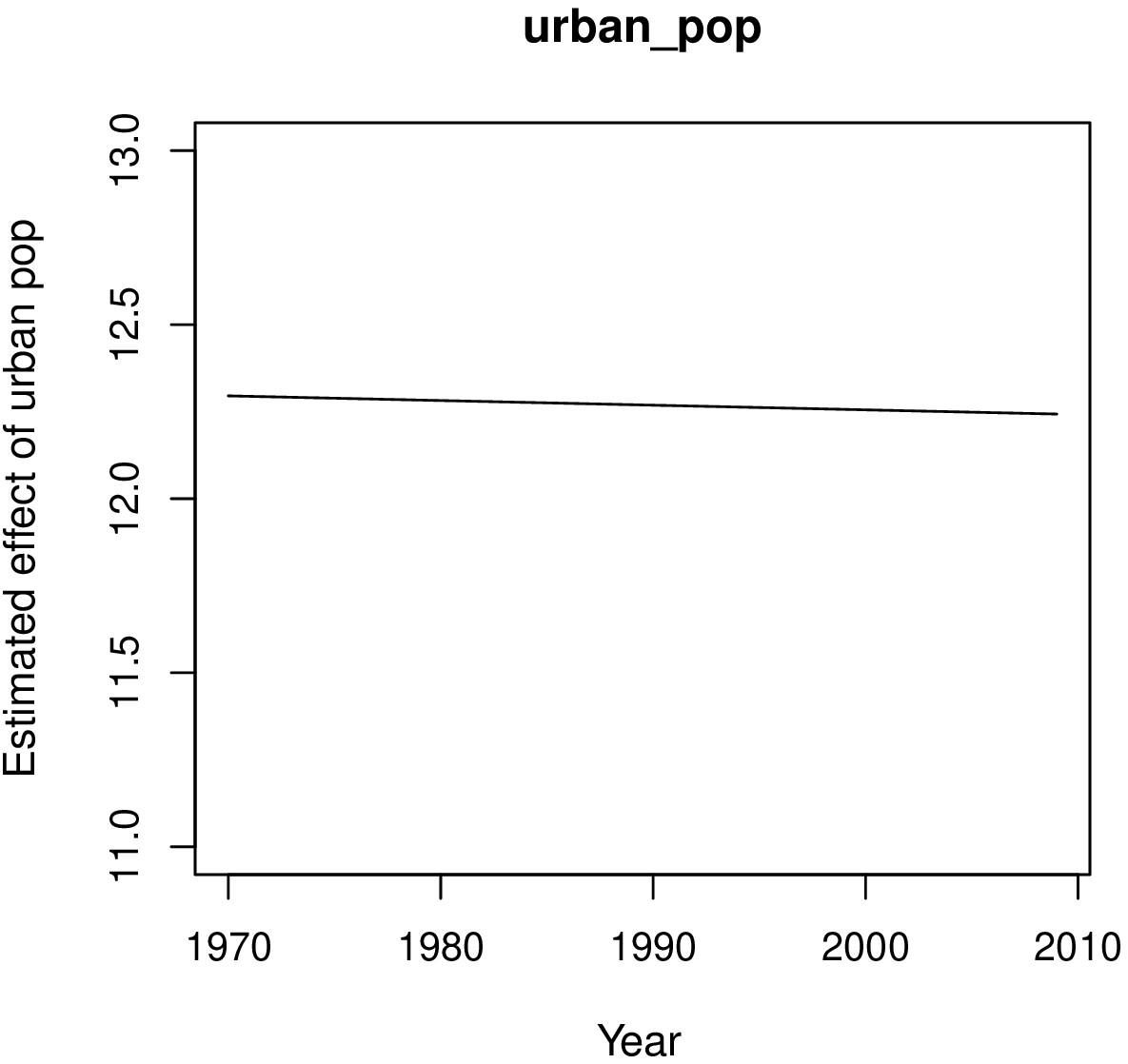}}\\
\end{tabular}
\end{center}
\caption{Estimate of the linear concurrent effect of GDP per capita (left panel) and urban population (right panel) on fisheries footprint (FSCAD).}
\label{fig:fig5}
\end{figure}

 We use the pre-whitening procedure discussed in Section \ref{sec:prewhite} and apply our proposed variable selection method. Out of 20 covariates, the proposed FSCAD and FMCP method both identify GDP per capita and urban population as the two significant predictors. Gross domestic product (GDP) is a measure of the market value of all the final goods and services produced in a specific time period, and GDP per capita is a measure of a country's economic output adjusting for its number of people. As the major economic indicator GDP per capita is associated with primary aspects of economic growth, consumer behaviour, trade and therefore is a key indicator of fisheries footprint of nations over time. Furthermore, GDP per capita is a common metric in extant social science research to operationalize the extent to which a nation is successfully developing according to the standards of the world, capitalist economy \citep{dietz2013structural}. 
 Figure \ref{fig:fig5} (left panel) shows the estimated regression curve for GDP per capita obtained by applying the FSCAD selection method. The estimate from the FMCP method is similar. We observe the net effect of GDP per capita on fisheries footprint to be positive and linear, although the magnitude of the effect has decreased over time.
 
Urban population being the key market, also plays a crucial role in the total seafood consumption of nations and therefore influences fisheries footprint. Change in urban population reflects urbanization and urbanization have important effects on food security and farming \citep{satterthwaite2010urbanization, cohen2010food}. Figure \ref{fig:fig5} (right panel) shows the estimated regression curve illustrating the effect of urban population on fisheries footprint. Here also we notice the net effect of urban population on fisheries footprint to be positive, although the effect appears to be more or less constant with a very marginal decrease (in log scale). 
 
Here, it is important to note that fisheries footprint represents the metabolic potential of an ecosystem to reproduce itself ecologically. According to the Food and Agriculture Organization of the United Nations \citep{FAO2016}, about 58 percent of global fish stocks are currently fully exploited, and about 55 percent of ocean territory (conservatively) was subjected to industrial fishing in the past year \citep{kroodsma2018tracking}. Thus, there is declining metabolic potential for the expansion of capture fisheries, which likely helps to account for why variable effects were stronger in earlier, more ecologically productive decades. 

\begin{figure}[H]
\centering
\includegraphics[width=.6\linewidth , height=.6\linewidth]{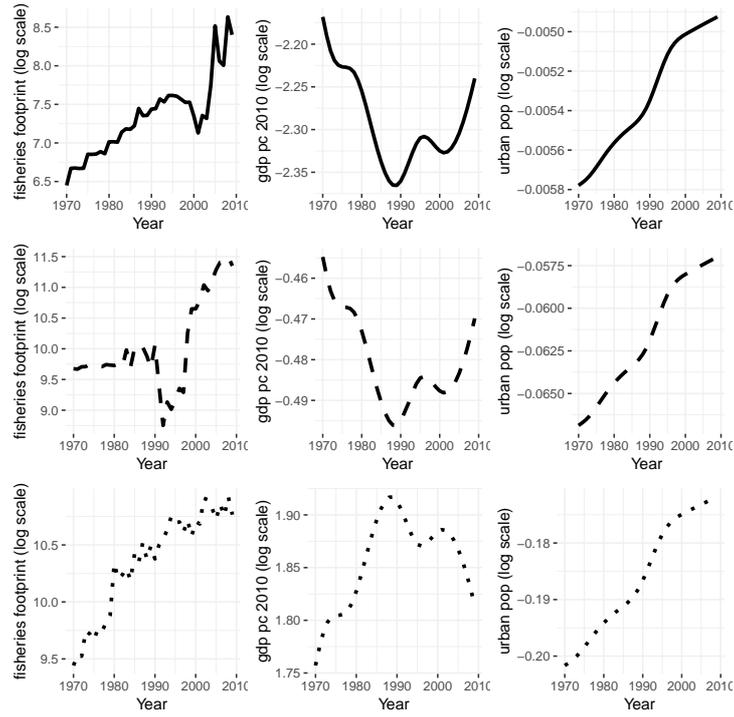}
\caption{Profile of  fisheries footprint, GDP per capita and urban population of three representative countries.}
\label{fig:fig7}
\end{figure}

Both these variables are important in the sense they represent the primary indicators in economics, food consumption, population dynamics, trade, etc; which directly interact with a nation's need for seafood and therefore should influence fisheries footprint. It is therefore not surprising that countries having high GDP per capita and/or high urban population e.g., United States, Australia, Singapore, etc also have a high fisheries footprint. In Figure \ref{fig:fig7} we display the fisheries footprint, GDP per capita and urban population profile of the three representative countries mentioned earlier. We notice the overall trend in the fisheries footprint profile can be described well by their GDP per capita and urban population profile, both of which were shown to have a positive effect on fisheries footprint.\\
\hspace*{- 5 mm}
\textit{Remark 3:}
We have successfully applied our proposed variable selection method to find out the relevant time-varying predictors and their time-varying effects on fisheries footprint. It is very plausible that there might be country or region specific effects on fisheries footprint as revealed in study by \cite{doi:10.1177/0020715219869976}, and one might be interested in estimating these effects. The proposed variable selection method for FLCM can be extended to handle such region specific effects in its existing form.

\hspace*{- 7 mm}
\textit{Remark 4:} 
We have considered concurrent effects of the predictors while some of the predictors might have lagged effects on fisheries footprint. For example in an economic crisis or recession, the predictors could very likely have reverberating impacts on development for a few years. As invested capital takes time to flow through the economy, considering such lagged effects would be interesting. We applied our variable selection method with lagged predictors present along with the original predictors (with lag window = 1, 3). We found out that for lag one, the proposed FMCP and FSCAD method select almost identical models with the FSCAD method selecting `services value growth pct' as an additional variable. Considering a lag window of three years, the FSCAD method selects urban population lag instead of urban population while the FMCP method additionally selects `aquaculture production tons' and `services value growth pct lag' as influential covariates. These results indicate some of the predictors could have reverberating impacts and a more general framework like the historical functional regression model \citep{malfait2003historical} might be more suitable to model past effects of covariates on the response at current time point.   
\section{Discussion}
\label{con}
In this article, we have proposed a variable selection method in functional linear concurrent regression extending the classically used penalized variable selection methods like LASSO, SCAD, and MCP. We have shown the problem can be addressed as a group LASSO and their natural extension group SCAD or group MCP problem. We have used a pre-whitening procedure to take into account the temporal dependence present within functions and through numerical simulations, have illustrated our proposed selection method with group SCAD or group MCP penalty can select the true underlying variables with high accuracy and has minuscule false positive and false negative rate even when data is observed sparsely, is contaminated with measurement error and the error process is highly non-stationary. We have illustrated usefulness of the proposed method by applying to two real datasets: the dietary calcium absorption study data and the fisheries footprint data in identification of the relevant time-varying covariates. In this article we have used a resampling subject based bootstrap method to measure uncertainty of the regression functions estimates, theoretical properties corresponding to such bootstrap is something we would like to explore more deeply in the future.

There are many interesting research directions this work can head into. In real data, the dynamic effects of the predictors might always not be linear. In future, we would like to extend our variable selection method to nonparametric functional concurrent regression model \citep{maity2017nonparametric}, which is a more general and flexible model to capture complex relationships present between the response and covariates. \hspace{- .8 mm}As mentioned earlier it would be also of interest to consider the lagged effects of covariates through a more general historical functional regression model \citep{malfait2003historical}. 

In developing our method we assumed the covariates to be independent and identically distributed. In many cases this might not be a reasonable assumption. For example, in the fisheries footprint data some countries could be very similar and form clusters, on the other hand, they might not be even independent with the interplay of economies and other variables among nations. Even if the covariates are not independent over subjects, the variable selection criterion proposed in this article can still be used in practice as a penalized least square method. The heterogeneity present among the subjects can be addressed using interaction effect of covariates with regions, which can be clustered based on the level of affluence. This can be done similarly as in \cite{doi:10.1177/0020715219869976}. Alternatively, one can also use subject specific functional random effects for covariates, especially if one is interested in individual specific trajectories. Functional linear mixed model \citep{liu2017estimating} might be an appropriate choice in such situations. Extending the proposed variable selection method to such general functional regression models would be an extension of this work and remain an area for future research.
  
\section*{Software} 
All the methods discussed in this article has been implemented using
the `grpreg' package \citep{breheny2019package} in R. Illustrations of implementation of our method using $R$ are available with this article on Wiley Online Library and at GitHub (\url{https://github.com/rahulfrodo/FLCM_Selection}).
\section*{Acknowledgement} 
We would like to thank the editor, the associate editor and an anonymous referee for their valuable inputs and suggestions which have greatly helped in improving this article.


\bibliographystyle{rss}
\bibliography{sel}
\end{document}